\documentclass[sigconf]{acmart}

\usepackage{amsmath,amssymb,amsfonts}
\usepackage{sansmath}
\usepackage{paralist}
\usepackage{csquotes}
\usepackage{textcomp}
\usepackage{booktabs}
\usepackage{tabularx}
\usepackage{multirow}
\usepackage{numprint}
\usepackage{enumitem}
\usepackage[capitalise]{cleveref} \Crefname{table}{Tab.}{Tabs.}
\usepackage{subcaption}  
\usepackage{graphicx} \graphicspath{{./Material/}}
\usepackage{balance}

\definecolor{color-purple}{RGB}{109,66,171}
\definecolor{color-green}{RGB}{146,193,68}
\definecolor{color-orange}{RGB}{241,143,35}
\definecolor{color-red}{RGB}{147,34,15}
\definecolor{color-blue}{RGB}{31,81,116}

\definecolor{color-gradient-1}{RGB}{147,34,15}
\definecolor{color-gradient-2}{RGB}{158, 85, 26}
\definecolor{color-gradient-3}{RGB}{169, 138, 38}
\definecolor{color-gradient-4}{RGB}{172, 181, 52}
\definecolor{color-gradient-5}{RGB}{146,193,68}
\newcommand{\eg}{e.g.\ }															% For example (e.g.)
\newcommand{\ie}{i.e.\ }															% That means (i.e.)
\newcommand{\eq}{\!=\!}																% Equal sign
\newcommand{\gt}{\!>\!}																% Greater sign

\newcommand{\perc}[1]{$#1$\,\%}														% Percentage

\newcommand{\Mit}{\textit{M}}														% Mean value
\newcommand{\SDit}{\textit{SD}}														% Standard deviation
\newcommand{\M}[1]{$\Mit \eq #1$}													% Mean value (with value)
\newcommand{\SD}[1]{$\SDit \eq #1$}													% Standard deviation (with value)
\newcommand{\MSD}[2]{\M{#1}, \SD{#2}}												% Mean value and standard deviation (M, SD)
\newcolumntype{Y}{>{\centering\arraybackslash}X} 									% New column type for centered columns
\newcolumntype{Z}{>{\raggedright\arraybackslash}X} 									% New column type for ragged-right columns

\setcopyright{rightsretained}
\acmDOI{}

\acmISBN{}

\acmConference[LBRS@RecSys '18]{Late-Breaking Results track part of the Twelfth ACM Conference on Recommender Systems (RecSys '18)}{October 2-7, 2018}{Vancouver, BC, Canada}
\acmBooktitle{Proceedings of the Late-Breaking Results track part of the Twelfth ACM Conference on Recommender Systems (RecSys '18), Vancouver, BC, Canada, October 2--7, 2018}
\acmYear{2018}
\copyrightyear{2018}

\acmArticle{}
\acmPrice{}

\begin{document}
\title{Understanding Latent Factors Using a GWAP}

\author{Johannes Kunkel, Benedikt Loepp, J\"urgen Ziegler}
\affiliation{%
  \institution{University of Duisburg-Essen, Duisburg, Germany}}
\email{{firstname.lastname}@uni-due.de}

\begin{abstract}
Recommender systems relying on latent factor models often appear as black boxes to their users. Semantic descriptions for the factors might help to mitigate this problem. Achieving this automatically is, however, a non-straightforward task due to the models' statistical nature. We present an output-agreement game that represents factors by means of sample items and motivates players to create such descriptions. A user study shows that the collected output actually reflects real-world characteristics of the factors.

\end{abstract}




\maketitle

\setlength{\textfloatsep}{2pt plus 0pt minus 0pt}
\setlength{\floatsep}{2pt plus 0pt minus 0pt}
\setlength{\intextsep}{2pt plus 0pt minus 0pt}
\setlength{\belowcaptionskip}{0pt}

\section{Introduction and Related Work}
\label{sec:introduction}

\emph{Recommender Systems} (RS) make it often difficult for users to understand the results, in particular when using model-based techniques such as \emph{Matrix Factorization} (MF) \cite{koren2015advances}: While latent factor models are known for accuracy and efficiency, they are typically considered non-transparent \cite{rossetti2013towards}. Some works indicate that learned factors are related to actual real-world characteristics \cite{koren2015advances, loepp2018interactive}, but only few steps have been taken to automatically explain the abstract dimensions, \eg by associating them with mined topics \cite{rossetti2013towards} or tags \cite{loepp2018interactive}. Others visualized relations between predefined tags and factors \cite{nemeth2013visualization} or reduced the dimensionality to display a map \cite{kunkel20173d}. Still, making factor meanings explicit can be considered difficult, especially without predefined data or complex visualizations. Thus, it seems promising to count on voluntary user contribution. \emph{Games with a Purpose} (GWAP), with their prominent method of \emph{Output-Agreement} (OA) \cite{Ahn.2008}, are well known for motivating users to solve such a human computation problem. In OA, randomly matched pairs of players are presented with a common input and have to come up with the same output without any means of communication \cite{Ahn.2008}. The winning strategy is thus to type in terms that describe the shared content as best as possible. Such games are often used to annotate images \cite{Ahn.2008}, but also for \eg eliciting preferences in RS \cite{Hacker.2009}.

In this paper, we present a GWAP that follows the OA method for collecting semantic descriptions of latent factors.
\section{The Game}
\label{sec:game}

In our game\footnote{Introduced in \cite{kunkel2018onlinespiel} in German language, see also: \url{http://interactivesystems.info/lfg/}.}, representative items are shown as shared input for one factor after the other. For producing the same output, players have to type in commonalities of these representatives. As descriptions, we collect all terms typed in for each factor. Such term-factor relations may then enable to \eg explain MF results by showing terms related to factors relevant for the current recommendation.

\emph{Factor Model and Representatives:} First, we use a \emph{Mahout ParallelSGDFactorizer} for offline model training. With \emph{MovieLens 20M} dataset, $10$ factors, $\lambda \eq 0.001$ and $16$ iterations, results were up to standard ($\mathit{RMSE} \eq 0.86$, $\mathit{NDCG@10} \eq 0.82$). While our approach is in principle independent of MF algorithm and parametrization, we deliberately choose a comparatively small number of factors. This is in line with earlier suggestions \cite{Loepp.2014} and allows us to a) collect more terms per factor with less effort, and b) decrease likelihood of factors being redundant, thus diversifying game experience.

Next, for selecting sets of representatives both distinguishable and reflecting the semantics of model dimensions, we basically follow \cite{Loepp.2014}: We calculate for each factor $f$ and each movie $i$ that has a value in the upper \perc{25} of values for $f$ a score $s_\mathit{if} \eq 0.4 \cdot \mathit{pop}(i) + 0.3 \cdot \mathit{rel}(i, f) + 0.3 \cdot \mathit{spec}(i, f)$, taking \emph{popularity} (high number of ratings), \emph{relevance} (high value for $f$), and \emph{specificity} (high value for $f$ but neutral for others) into account. For each factor, we then select the $25$ movies with highest $s_\mathit{if}$ as representatives. Weights and numbers are the result of pretesting.

\begin{figure}[ht!]
	\centering
	\setlength{\fboxsep}{0pt}%
	\setlength{\fboxrule}{0.5pt}%
	\fcolorbox{lightgray}{white}{\includegraphics[width=0.7\columnwidth]{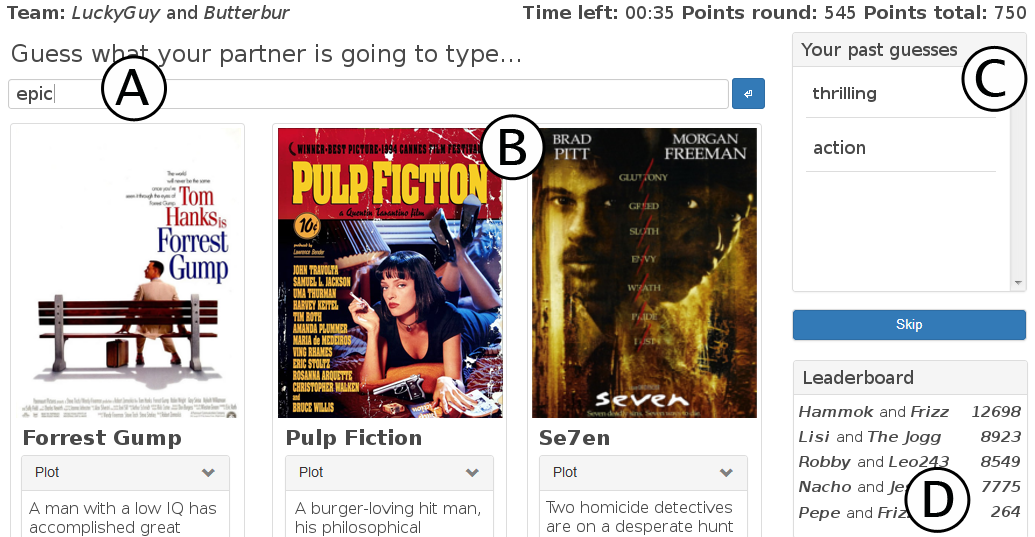}}
	\caption{\emph{LuckyGuy} guesses (A) how his partner would describe the factor representatives (B). He made two guesses already (C). A leaderboard shows overall performances (D).}
	\label{fig:screenshot}
\end{figure}

\emph{Game Mechanics:} We implemented the game as a web application (\cref{fig:screenshot}), as recommended for OA \cite{Ahn.2008}. A \emph{game} lasts $3$ min., during which two randomly matched players seek to play as many \emph{rounds} as possible with the goal of gaining \emph{points} and scoring high in a \emph{leaderboard}. Rounds are randomly related to factors of the underlying MF model. In each round, $3$ movies are randomly chosen from the $25$ factor representatives, and displayed by means of poster, plot description, cast and director. A round ends a) as soon as a \emph{match} is found in the terms entered by both players, \ie they \emph{guess} the same, or b) if both decide to skip \eg because the movies are too hard to describe (leading to penalty points). Either way, they then proceed to the next round, \ie an item set for another factor is shown.

\begin{table*}[ht!]
	\setlength{\tabcolsep}{1pt}
	\setlength{\aboverulesep}{0pt}
	\setlength{\belowrulesep}{1pt}
	\centering
	\caption{Latent factors with sample representatives, number of guesses and matches, and terms that led to at least two matches.}
	\label{tab:term-overview}
	{\tiny{
	\begin{tabularx}{\textwidth}{lZZZZZZZZZZ}
		\toprule
			\textbf{Factor} & \textbf{1} 	& \textbf{2} 	& \textbf{3} 	& \textbf{4} 	& \textbf{5} 	& \textbf{6} 	& \textbf{7} 	& \textbf{8} 	&	\textbf{9} 	& \textbf{10} \\
		\midrule
			\parbox[t]{1.2cm}{\textbf{Sample \\ representatives}}
							& \emph{The Lion King}, \emph{Bad Boys}, \emph{Home Alone}
											& \emph{Forrest Gump}, \emph{Fight Club}, \emph{Se7en}
															& \emph{Jurassic Park}, \emph{Rocky V}, \emph{Star Wars}
																			& \emph{Cast Away}, \emph{Meet the Parents}, \emph{Waterworld}
																							& \emph{The Doors}, \emph{The Beach}, \emph{Casino}
																											& \emph{Indiana Jones}, \emph{Speed}, \emph{Aliens}
																															& \emph{Nude Girls}, \emph{American Pie}, \emph{Harold \& Kumar}
																																			& \emph{The Net}, \emph{Dave}, \emph{Groundhog Day}
																																							& \emph{Batman Forever}, \emph{Deep Impact}, \emph{Twister}
																																											& \emph{Pretty Woman}, \emph{Big}, \emph{E.T.}  \\
		\midrule
			\textbf{Guess./match.\ (ratio)}
							& 620 / 54 (11.48)
											& 612 / 57 (10.74)
															& 589 / 51 (11.55)
																			& 565 / 49 (11.53)
																							& 562 / 55 (10.22)
																											& 580 / 65 (8.92)
																															& 423 / 42 (10.07)
																																			& 438 / 50 (8.76)
																																							& 754 / 66 (11.42)
																																											& 598 / 56 (10.68)  \\
		\midrule
			\textbf{Matches (\#)}
							& \parbox[t]{2cm}{comedy (10) \\ funny (8) \\ disney (4) \\ action, love (3) \\ fight, sex (2)}
											& \parbox[t]{2cm}{action (13) \\ fight, man, serious (4) \\ thrilling (3) \\ comedy, dog, drama, \\ \-\hspace{.33em} thriller, war (2)}
															& \parbox[t]{2cm}{action (12) \\ war (5) \\ fight (4) \\ comedy (3) \\ drama (2)}
																			& \parbox[t]{2cm}{action (13) \\ comedy (8) \\ drama, funny, \\ \-\hspace{.33em} spooky, thriller (2)}
																							& \parbox[t]{2cm}{action (7) \\ comedy, horror, \\ \-\hspace{.33em} love (5) \\ spooky (3) \\ erotic, mystery, \\ \-\hspace{.33em} old (2)}
																											& \parbox[t]{2cm}{action (24) \\ comedy (3) \\ adventure, alien, \\ \-\hspace{.33em} fight, love, \\ \-\hspace{.33em} old, weapons (2)}
																															& \parbox[t]{2cm}{comedy (10) \\ sex (7) \\ action, college, \\ \-\hspace{.33em} drama (2)}
																																			& \parbox[t]{2cm}{love (12) \\ comedy (5) \\ family (3) \\ action, america, \\ \-\hspace{.33em} boring, romance, \\ \-\hspace{.33em} sex, woman (2)}
																																							& \parbox[t]{2cm}{action (18) \\ horror (6) \\ sci-fi, spooky (3) \\ alien, aliens, \\ \-\hspace{.33em} batman, comedy, \\ \-\hspace{.33em} future (2)}
																																											& \parbox[t]{2cm}{love (11) \\ comedy (6) \\ family (4) \\ action, animals, \\ \-\hspace{.33em} romance, romantic (3) \\ dramatic (2)}\\
		\bottomrule
	\end{tabularx}}}
	\vspace{-0.33cm}
\end{table*}
\section{Evaluation}
\label{sec:evaluation}

We conducted a user study to evaluate subjective game experience and collect a first baseline of factor descriptions. We recruited $84$ ($42$ female) participants (age: \MSD{21.23}{4.53}), which were asked to play the game and fill in a questionnaire. The study took place at our lab, at participants' home and in university classes. A supervisor was present and controlled that no communication occurred. Yet, few participants played again later without being supervised. Due to the different settings, players sometimes knew each other very well (\perc{21}), while in many other cases, they did not (\perc{42}). We used self-generated items to assess game-specific aspects, elicited demographics and domain knowledge, and measured user experience (SUS \cite{brooke1996sus}) and enjoyment (IMI \cite{Ryan.1982}). If not stated otherwise, we used 5-point Likert scales. We logged all interaction data, especially guesses and matches, to analyze the produced output.

\emph{Results:} $173$ games were played. Participants were required to play only once, but each player played on average $4.12$ games. Together with the mean score of $4.79$ (\SD{{1.26}}) on the 7-point enjoyment subscale of IMI, this indicates that they liked playing. Participants reported that they were only sometimes in doubt when entering guesses (\MSD{2.75}{1.05}) and found the game overall rather easy to play (\MSD{3.21}{1.24}). They tended to love movies (\MSD{3.31}{1.03}) and had average knowledge about recent ones (\MSD{2.60}{0.98}). Accordingly, representatives were to some extent known (\MSD{2.77}{0.99}), but it was pointed out that a few old movies should have been omitted. Still, participants reported to have somehow understood why movies were displayed together (\MSD{2.74}{0.96}), and that they, considering all rounds, seemed diverse (\MSD{3.58}{1.01}). Provided information appeared sufficient (\MSD{3.06}{0.95}), with posters (\MSD{4.32}{0.91}) being most informative. Usability was \emph{good} (SUS-score of $79$).

In total, $5\,741$ guesses were made, on average $574.10$ per factor (\SD{93.29}) and $33.18$ per game (\SD{20.60}). This resulted in a total of $545$ matches, on average $54.50$ per factor (\SD{7.23}) and $3.15$ per game (\SD{5.05}). Thus, each player had an \emph{expected contribution} \cite{Ahn.2008} of $68.35$ guesses and $6.49$ matches. For further analysis, we cleaned the dataset and set $X \eq 2$ as \emph{good label threshold} \cite{Ahn.2008}, \ie minimum number of matches for a term to be considered meaningful. This left us with $325$ matches comprising $35$ distinct terms. \cref{tab:term-overview} shows the collected data. Based on these terms, we created a dictionary and calculated content vectors by means of \emph{TF-IDF}, representing how often terms led to a match for a factor in relation to how often this was overall the case. This allowed us to compare the sets of terms, \ie descriptions created for factors, by means of vector cosine similarity. Overall, similarities were very low (\MSD{0.09}{0.13}).

\emph{Discussion:} Questionnaire results and cosine similarities indicate high diversity between factors. Apparently, the method for selecting representatives ensures that shown items reflect different factor semantics. Accordingly, differences can be found in the sets of collected terms: Matches and their frequency seem consistent within factors, but vary between (see \cref{tab:term-overview}). Only in few cases, factors seem less unique, \eg $8$ and $10$ ($\mathit{sim} \eq 0.54$). This could be due to insufficient data, making items less distinguishable. On the other hand, factors might actually express similar aspects.

Some factors appear to have more obvious semantics: For instance, guess-match ratios in \cref{tab:term-overview} show that players arrived at a match more often for factors $6$ and $8$, \ie they are easier to describe.

Overall, current game mechanics seem to favor rather general terms, \eg genres such as \enquote{action} or \enquote{comedy}. This effect has also been shown earlier \cite{Ahn.2008}, and can be prevented \eg by taboo lists. However, implementing such mechanics would have required output data which we had not had prior to our study. On the other hand, very specific terms led to matches as well, \eg \enquote{dog} for factor $2$. This does not appear to result from participants describing a factor's general meaning, but from certain movie posters being displayed. Yet, as the descriptions are only affected to a small degree, this issue will most likely vanish with more output data and $X \gt 2$.
\section{Conclusions and Outlook}
\label{sec:conclusions-outlook}

Study results show that our GWAP is fun to play and thus motivates users to produce output useful to better understand the hidden semantics of common RS models: Terms entered by players allow deriving meaningful and distinguishable latent factor descriptions, which may be used \eg to explain recommendations by presenting keywords related to factors relevant for the active user and the respective items. Yet, investigating application areas is subject of future work, as is implementing a single player version as well as advanced game mechanics such as taboo lists for collecting further output data and more specific terms via gameplay.

\balance

\bibliographystyle{ACM-Reference-Format}
\bibliography{Bibliography}

\end{document}